\documentclass[tighten,apjl]{emulateapj}
\usepackage[]{times, graphicx}
\newcommand{\soho}{{\em SOHO{}}}

\newcommand{\hinode}{{\em Hinode{}}}
\newcommand{\stereo}{{\em STEREO{}}}
\newcommand{\sdo}{{\em SDO{}}}
\newcommand{\pref}{\protect\ref}
\newcommand{\ace}{{\em ACE{}}}
\newcommand{\wind}{{\em WIND{}}}
\newcommand{\uly}{{\em Ulysses{}}}
\newcommand{\dfe}{$\mathrm{D}_{{\mathrm{Fe}}}$}
\newcommand{\qfe}{$\langle \mathrm{Q}_{{\mathrm{Fe}}}\rangle$}
\newcommand{\vsw}{$\mathrm{V}_{{\mathrm{SW}}}$}
\newcommand{\ahe}{$\mathrm{A}_{{\mathrm{He}}}$}

\begin{document}

\shorttitle{Variations of Plasma Heating in the Fast Solar Wind}
\shortauthors{S.~W. McIntosh et al.}
\title{Solar Cycle Variations in the Elemental Abundance of Helium and Fractionation of Iron in the Fast Solar Wind \-- Indicators of an Evolving Energetic Release of Mass from the Lower Solar Atmosphere}
\author{Scott W. McIntosh\altaffilmark{1}, Kandace K. Kiefer\altaffilmark{2,1}, Robert J. Leamon\altaffilmark{3}, 
Justin C. Kasper\altaffilmark{4}, Michael S. Stevens\altaffilmark{4}} 

\altaffiltext{1}{High Altitude Observatory, National Center for Atmospheric Research, P.O. Box 3000, Boulder, CO 80307}
\altaffiltext{2}{Department of Earth \& Atmospheric Sciences, Purdue University, West Lafayette, IN 47906}
\altaffiltext{3}{Department of Physics, Montana State University, Bozeman, MT 59717}
\altaffiltext{4}{Harvard-Smithsonian Center for Astrophysics, Smithsonian Astrophysical Observatory, Cambridge, MA 02138}

\begin{abstract}
We present and discuss the strong correspondence between evolution of the emission length scale in the lower transition region and in situ measurements of the fast solar wind composition during this most recent solar minimum. We combine recent analyses demonstrating the variance in the (supergranular) network emission length scale measured by \soho{} (and \stereo{}) with that of the Helium abundance (from \wind) and the degree of Iron fractionation in the solar wind (from the \ace{} and \uly{}). The net picture developing is one where a decrease in the Helium abundance and the degree of Iron fractionation (approaching values expected of the photosphere) in the fast wind indicate a significant change in the process loading material into the fast solar wind during the recent solar minimum. This result is compounded by a study of the Helium abundance during the space age using the NASA OMNI database which shows a slowly decaying amount of Helium being driven into the heliosphere over the course of the several solar cycles.
\end{abstract}


\section{Introduction}
Decrypting the physical processes encoded in the compositional measurements of the solar wind is akin to the use of DNA analysis of modern forensic science or archaeology to piece together a picture of what happened to whom, by who, when, and where. The streams of particles captured by the compliment of solar wind composition instruments in interplanetary space are a literal treasure trove of physical information about the plasma from the time that it was originally heated in the solar atmosphere, through any interactions it had with other streams en route to the detector. The distribution of ion charge states, densities, speeds, and atomic abundances among other things measured by these instruments form the DNA of the outer solar atmosphere's energetic processes and provide a significant challenge to our understanding as we are limited, for now, to remote sensing its day to day behavior and subtle evolution over decades.

\begin{figure}
\epsscale{1.15}
\plotone{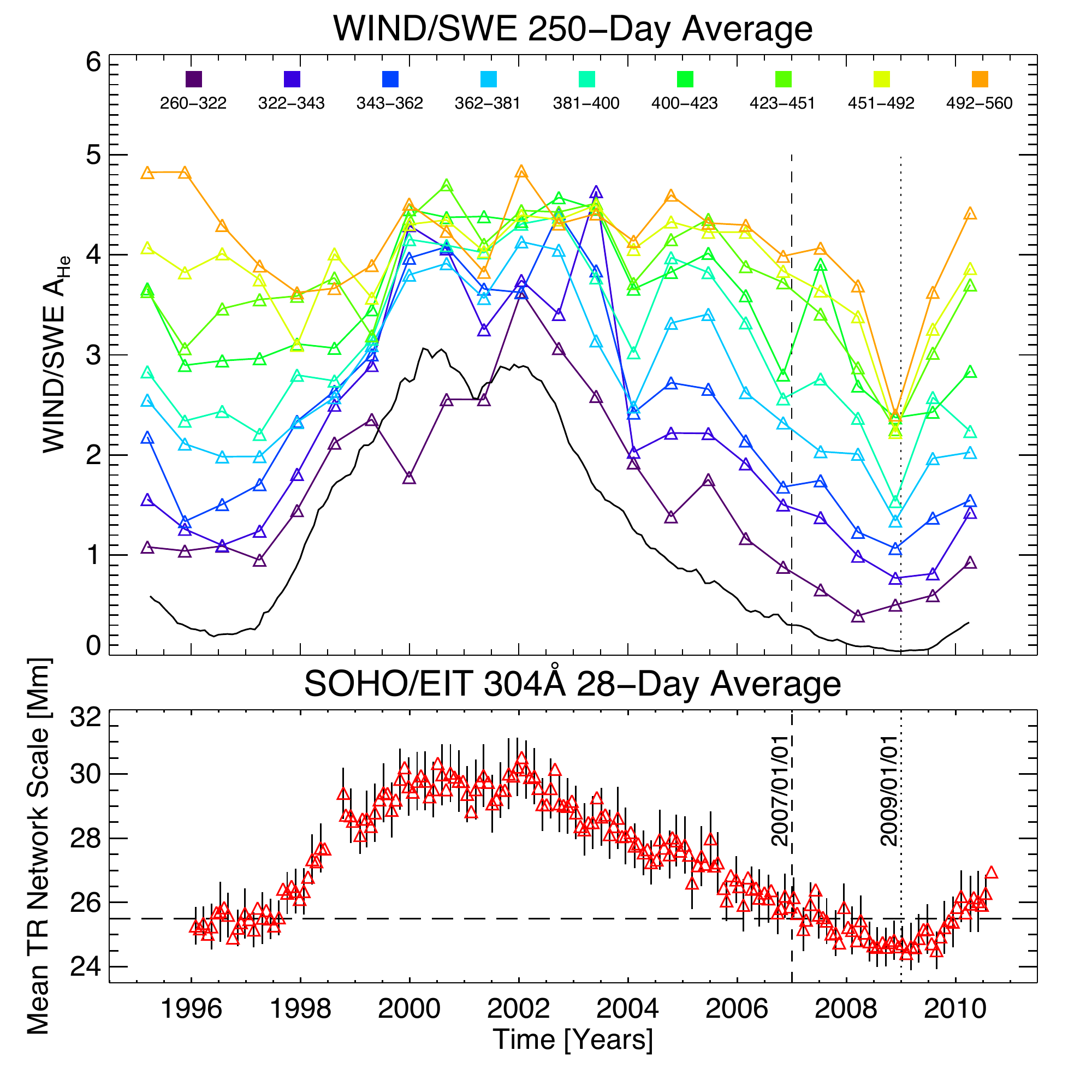}
\caption{Combining the analysis of the \wind{} Helium abundance (\ahe) as a function of the solar wind speed in 250-day averages in the top panel \citep[from][]{Kasper2011} with the 28-day running average of the transition region (supergranular) network length scale from \soho{} \citep[from][]{2011ApJ...730L...3M}. The solid black trace in the top panel shows the variation in the smoothed monthly sunspot number over the same time period. The dashed vertical line marks January 1 2007 as an approximate date when the network scale falls below the mean value of the 1996 solar minimum. The time of minimum scale and Helium abundance occur coincidently close to January 1 2009. \label{f1}}
\end{figure}

In this Letter we start an investigation into some of the compositional measurements performed through the long, and unanticipated, activity minimum between solar cycles 23 and 24 where the reduced magnetic activity of the star \citep[e.g.,][]{2008GeoRL..3522103S, 2009ApJ...707.1372W} has allowed us to study the most basic, or basal, of energetic processes occurring in the solar atmosphere and their impact on the heliospheric system \citep[e.g.,][]{2008GeoRL..3518103M,2009JGRA..11409105G,2010GeoRL..3716103S,2010GeoRL..3718101M,2010ApJ...723L...1M}. This investigation was motivated by results presented in two recent papers by \citet{2011ApJ...730L...3M} and \citet{Kasper2011}, where their key results are pictorially combined in Fig.~\pref{f1}.

\citet{2011ApJ...730L...3M} demonstrated a variance in the dominant scale length of emission in the low transition region (TR), the (supergranular) magnetic network through the cycle~23/24 minimum. It was proposed that the scale length change was driven by the network vertices (and boundaries to a lesser degree) of the TR network emission being  more sparsely packed with magnetic flux elements. The magnetic network reflects the scale length over which mass and energy are transported into the quiescent solar atmosphere \citep[quiet sun and coronal hole plasma alike;][]{McIntosh2007,2009ApJ...707..524M} and so it was anticipated that a reduction in the network scale, driven by a change in the underlying magnetism, would impact the amount, and possibly the temperature, of material being inserted into the corona and solar wind. During the time that the mean network scale dropped below the 1996 minimum value (indicated on the plot as a horizontal dashed line) \citet{Kasper2011} noted a profound drop in the Helium abundance \citep[\ahe{}; defined as 100 times the ratio of the measured alpha particle and proton densities][]{Aellig2001} of the fast solar wind measured by the \wind{} spacecraft below that of the previous solar minimum. Note that the slower wind streams show far more variance with the solar activity cycle as well as showing a drop in Helium abundance between the 1996 and 2009 minima. \citet{1996AIPC..382..269Y} identified a similar pattern of Helium abundance variability in solar wind streams with different magnetic (and atmospheric, e.g., coronal hole versus CME, etc) origins using Prognoz~7 observations. However, we choose to focus our thoughts on the fast wind abundance change, motivated by the relatively simplicity of fast wind origins in coronal holes \citep[e.g.,][]{McIntosh2010}, and also because its composition was considered to be relatively invariant \-- invariant to a degree that such a distinct, large amplitude, change is extremely curious.

The remainder of this Letter follows the premise discussed by \citet{Parker1991}, and later promoted by \citet{McIntosh2010} and \citet{McIntosh2011b}, that the fast solar wind is populated and accelerated by a tandem process. The first of these processes heating the material out of the chromosphere to temperatures above the 1MK \citep{DePontieu2011,2011ApJ...736....9M} necessary for the plasma to have enough thermal pressure to overcome gravity, accompanied by a second process that is able to accelerate the plasma to its final measured speed, e.g., low frequency Alfv\'{e}n waves \citep{DePontieu2007b,McIntosh2011b}. We have observed from Fig.~\pref{f1} that something has affected the deposition of Helium in the solar wind during the last solar minimum \-- we presume that the coincident reduction in scale of the magnetic network has resulted in the abundance change observed in the fast solar wind. In the following sections we explore some of the compositional measurements obtained by the SWICS instrument on the \uly{} \citep{1992A&AS...92..267G} and the SWICS/SWIMS package on \ace{} \citep{GloecklerEA98_short}  to explore the composition of the fast solar wind during the recent solar minimum in and out of the ecliptic plane. To complete this preliminary analysis we perform the analysis of \citet{Kasper2011} on the NASA SPDF OMNI solar wind database which provides alpha particle and proton densities back to the early 1970s, three complete solar cycles ago.

\section{Compositional Changes}
As we mentioned above, the composition of the solar wind likely reflects the physical conditions that the plasma meets from its original removal from the lower solar atmosphere until it finally finds itself on an open magnetic field line and travels outward into the heliosphere. Under our presumption above the fast solar wind composition then reflects the conditions set up by the rapid heating of the plasma from the chromosphere as illustrated recently in joint \sdo{}/\hinode{} observations \cite{DePontieu2011}. The slow solar wind, on the other hand, experiences some finite (but unknown) period of time in the closed corona of the quiet sun or active regions before being able to escape. Therefore, its composition is likely to be far more complex and variable \citep{Zurbuchen2002} as the phases of heating, cooling, ionization, and recombination are likely to play a role in setting up the mixture of ions and charge states that are observed in situ \citep[see also the discussion and cartoon presented in][]{DePontieu2009}. As we have said above, we will focus on the apparently simpler state of the fast solar wind.

In the following subsections we investigate the properties of ionic Iron measured in the fast solar wind by the \uly{} and \ace{} spacecraft \citep[see, e.g.,][]{vonSteiger2010}. We consider the degree of Iron fractionation, \dfe, which we define as the relative change in the measured Iron to Oxygen ratio (Fe/O) divided by the expected value of that ratio in the photosphere \citep[e.g., the figures of][]{GeissEA95}, or 0.035 \citep{2007SSRv..130..105G}.

\begin{figure}
\epsscale{1.15}
\plottwo{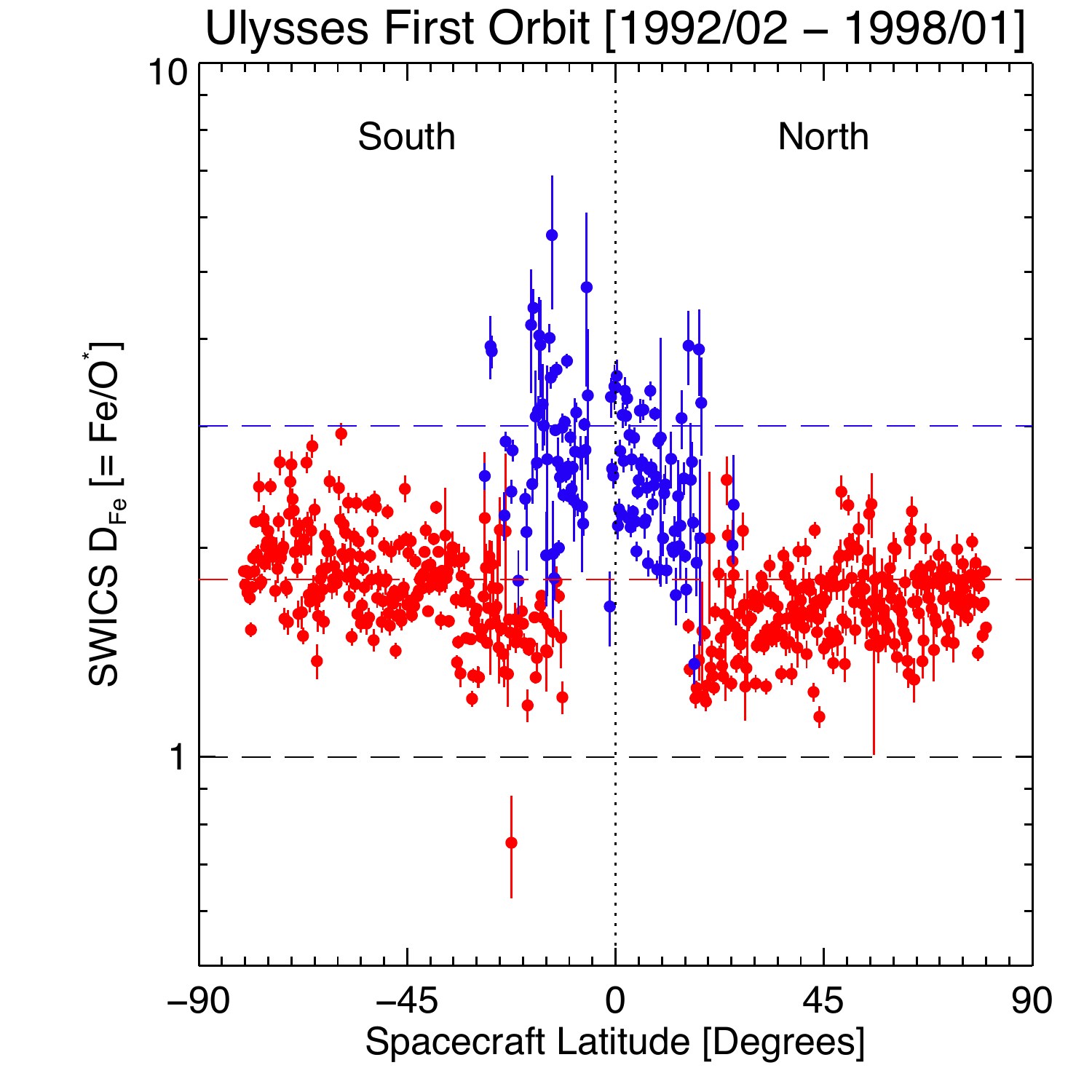}{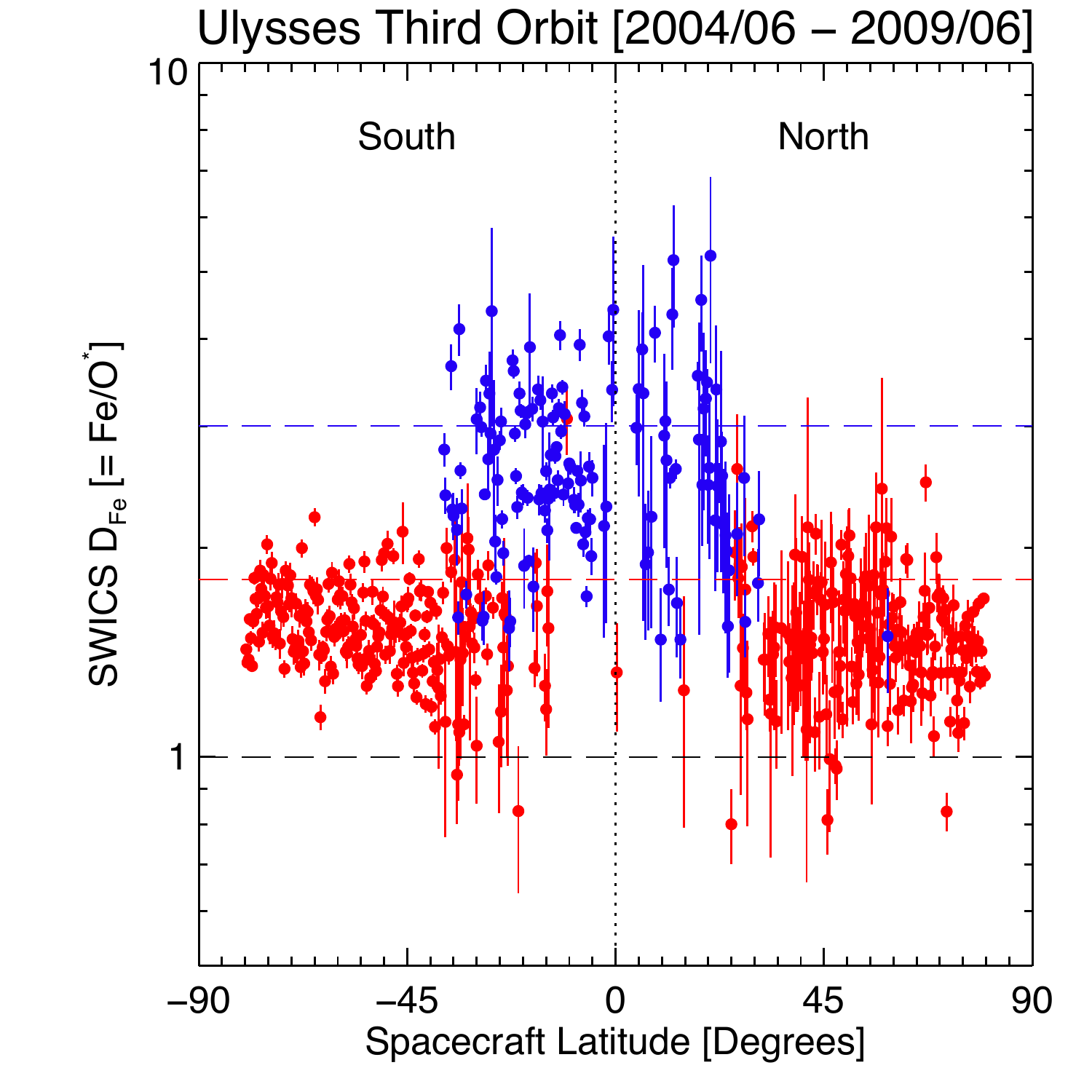}
\caption{Comparing the degree of Iron fractionation (\dfe, the ratio of the Iron and Oxygen densities normalized by the expected photospheric value 0.035 measured over the first (left) and third (right) orbits of \uly{} by SWICS as a function of the heliographic latitude of the spacecraft. These orbits spanned the minima of cycle~22 into 23 and that of cycle~23 into 24 respectively. We have isolated the fast (\vsw$>$500km/s; red) and slow (\vsw$<$400km/s; blue) and plotted in 0.5 degree bins of latitude. The horizontal dashed lines indicate values of \dfe{}$=1$ (black), \dfe{}$=1.8$ (red), and \dfe{}$=3$ (blue). \label{f2}}
\end{figure}

\begin{figure*}
\epsscale{1.15}
\plotone{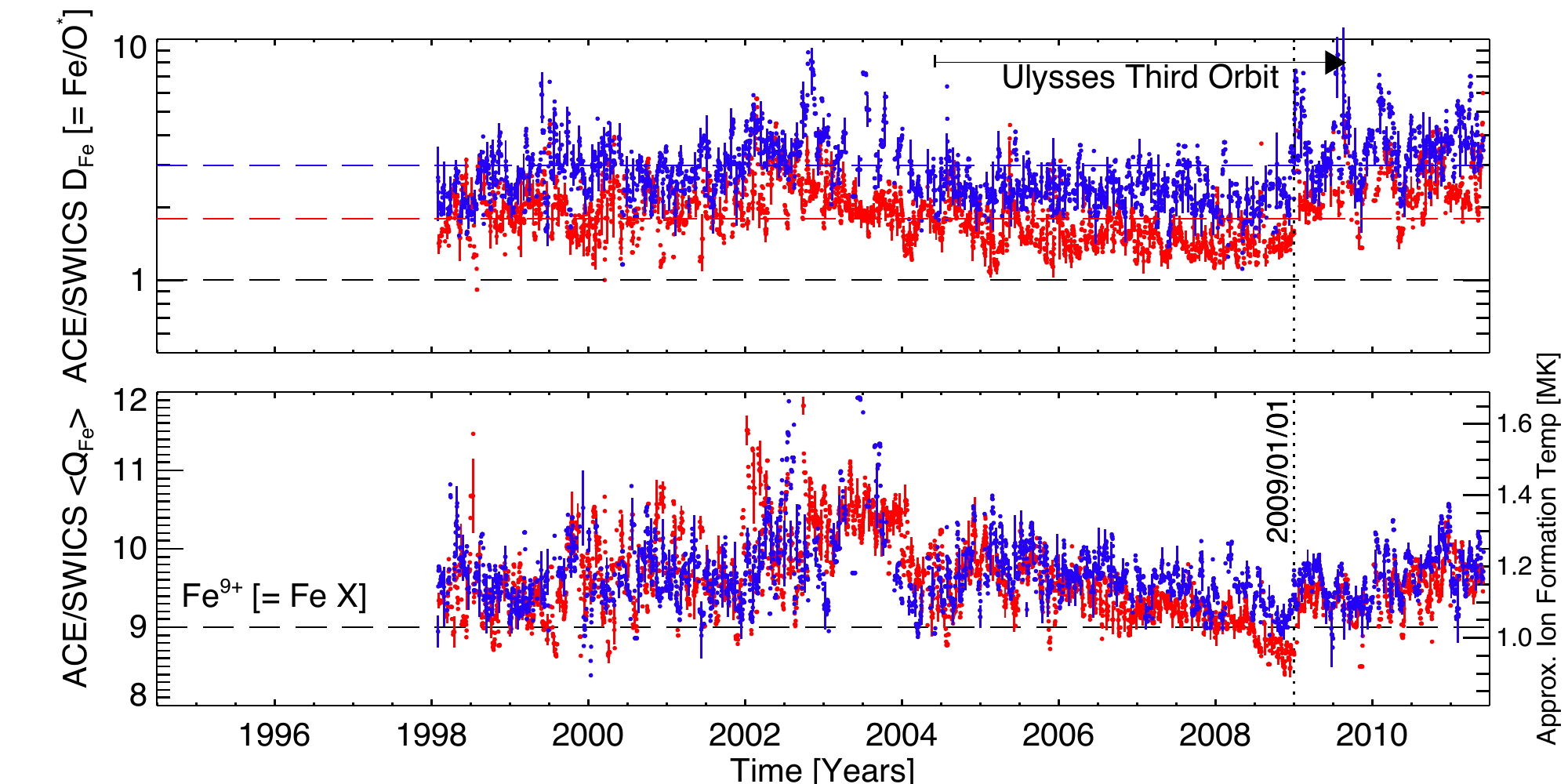}
\caption{Comparing the daily averages in the degree of Iron fractionation (\dfe; top) and the average Iron charge state (\qfe; bottom) in the fast (\vsw$>$500km/s; red) and slow winds (\vsw$<$400km/s; blue) through the 2009 solar minimum as measured by \ace{}/SWICS in the ecliptic plane \-- January 1 2009 is indicated by a vertical dotted line. In the upper panel the horizontal dashed lines indicate values of \dfe{}$=1$ (black), \dfe{}$=1.8$ (red), and \dfe{}$=3$ (blue) and an arrow is drawn to illustrate the duration of the third orbit of \uly{} through the declining phase of cycle~23. In the bottom panel a horizontal dashed line is drawn at \qfe$=9$ for the reference of the reader. \label{f3}}
\end{figure*}

\subsection{Ulysses/SWICS \dfe}
Fig.~\pref{f2} shows the difference between \dfe{} in the fast and slow solar wind between the two solar ``minimum'' orbits of \uly{}, its first (left; 1992--1998) and third (right; 2004--2009), as a function of the spacecrafts heliospheric latitude\footnote{Orbit one covered the declining phase of cycle~22 and ascent of cycle~23 while orbit three sampled the declining phase of cycle~23 into and through the 2009 minimum.}. The points and bars in the plot reflect the mean and standard deviations of \dfe{} averaged over 0.25 degree latitude bins for wind speeds less than 400km/s (slow wind; blue) and greater than 500km/s (fast wind; red). The horizontal dashed lines show values of \dfe{}$=1$ (black), \dfe{}$=1.8$ (red), and \dfe{}$=3$ (blue). The latter two are representative of typical fast and slow wind degrees of Iron fractionation \citep[see, e.g., Fig. 2 of][]{GeissEA95}. Comparing the left and right panels of the figure we see that the measured values of \dfe{} are systemically lower in the third orbit of \uly{} compared to that of its first.

We have repeated the analyses shown in Fig.~\pref{f2} using the technique of \cite{vonSteiger2010} that discriminates between ``fast'' and ``slow'' wind based on charge state ratios rather than speed. However, the results are qualitatively (if not quantitatively) the same as those shown here, and we shall only further discuss solar wind speed as the discriminator.

\subsection{ACE/SWICS \dfe{} and Average Iron Charge}
Fig.~\pref{f3} shows the variation of \dfe{} in the top panel and the average Iron charge state, \qfe{} in the bottom panel over the entire \ace{} mission \-- unfortunately \ace{} did not sample the 1996 minimum, but we will use the observations to look for systematic changes in the declining phase of cycle~23 and the 2009 solar minimum in the ecliptic plane. Isolating the fast and slow wind streams in the same way as above (red for \vsw$>$500km/s; blue for \vsw$<$400km/s) and averaging the measurements of Fe/O (and \qfe) over a day we see the systemic difference between fast and slow wind values of \dfe{} where we have again drawn the dashed lines indicating values of 1, 1.8, and 3. The fast and slow winds show a systematically offset slow decline in \dfe{} in the declining phase of cycle~23, with the latter reaching values of the order 1.3 before rapidly increasing again at the start of 2009. The lower panel of Fig.~\pref{f3} shows the variance of  \qfe{} for fast and slow wind streams through the cycle~23 solar maximum, through the declining phase and into the deep solar minimum of 2009. The values of \qfe{} in the fast and slow wind drop from $10$ in 2005 to a value of $\sim$8.5 for the fast wind and $\sim$9 for the slow wind at the start of 2009 before increasing rapidly. Note the dashed line drawn for Fe$^{9+}$ (or ``\ion{Fe}{10}'' in spectroscopic notation) and that the right hand scale of the lower panel shows an approximate value for the temperature at which that charge state reaches maximum population, roughly $T \sim 10^4 (Q+1)^2$~K. So, not only does the value of \dfe{} drop during the 2009 solar minimum to values consistent with the measurements of \uly{} out of the ecliptic plane, but the inferred temperature of the plasma (under equilibrium conditions) appears to be systematically lower into the period where the network scale reached its minimum.

\section{Steady Decay Over Many Minima?}
\begin{figure*}
\epsscale{1.15}
\plotone{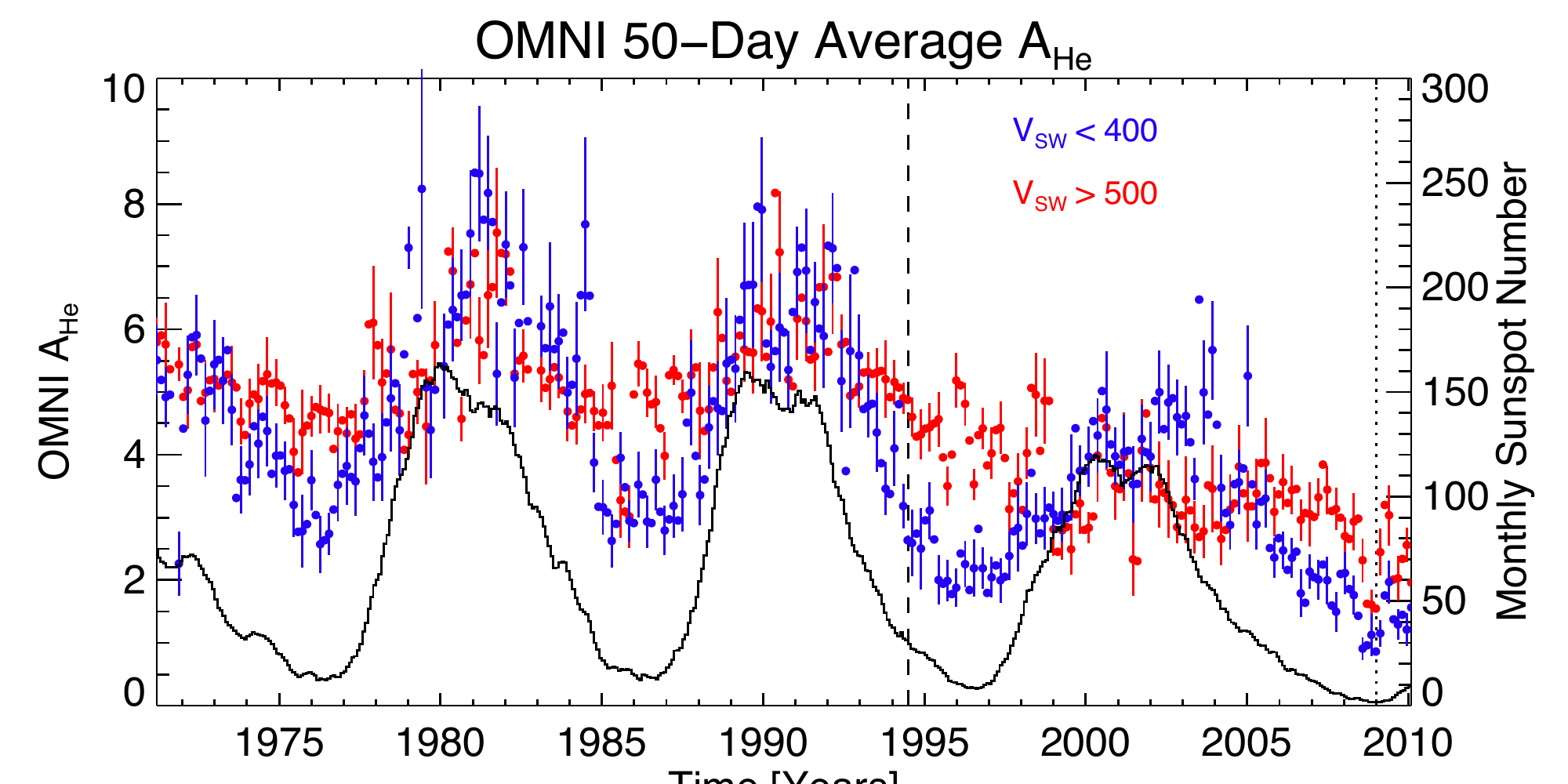}
\caption{Using the OMNI database to extend 50-day averages of \ahe{} shown in Fig.~\pref{f1} back through other cycles for fast (\vsw$>$500km/s; red) and slow (\vsw$<$400km/s; blue) winds. Again, January 1,~2009 is indicated by a vertical dotted line while June 1,~2004 is indicated by a vertical dashed line to mark the start of the \wind{} measurements shown in Fig.~\pref{f1}. The solid black trace shows the variation in the smoothed monthly sunspot number over the time period.  \label{f4}}
\end{figure*}

In Fig.~\pref{f4} we show the results of performing the velocity-discriminating \ahe{} analysis of \citet{Kasper2011} on the NASA SPDF OMNI solar wind database record of the alpha/proton ratio. Separating the fast ($>$500km/s; red) and slow ($<$400km/s; blue) wind streams and performing a 50-day running average we can compare the variation in the solar wind \ahe{} with magnetic activity back to 1970, noting that from 1994 onward many of the measurements come from \wind{}, like those in Fig~\pref{f1}. Like Fig.~\pref{f1} there is stronger modulation in \ahe{} in the slow wind, but is still present in the fast wind, which also shows a systematically slower decay in the descending phase of the cycles. Over the entire time period, and particularly from 1980, the values of \ahe{} inferred show a steady decline with the reduction in large scale solar activity. While this plot is striking in its trends and variance, it must be considered carefully as it is a unique single record composed of measurements from many spacecraft and, as such, is dependent on the inter-calibration of those measurements, but since there is no obvious sign of discontinuous jumps in the record over with a range of averaging windows it would appear that the OMNI alpha/proton record is of high internal consistency. The measure of this variance is the continued monitoring of these values in the ecliptic plane into and through the ascending phase of cycle~24.

\section{Discussion}
We have observed a significant decrease in the Helium abundance in the fast solar wind occur in the just past solar minimum. In addition, we have observed that the degree of Iron fractionation measured by the SWICS instruments on the \uly{} and \ace{} spacecraft have receded from their typical value of $\sim$2 and have approached values nearer unity over the same period in time. The latter spacecraft has also allowed us to investigate the variation in the mean charge state of Iron and a clear, steady drop is seen in the fast wind over the declining phase of solar cycle~23 hitting a minimum value at the start of 2009. 

On the Sun at this time, the (supergranular) network length scale reached a minimum. That length scale, mediated by the distribution of the small-scale photospheric magnetic field, dictates the flow of mass and energy into the quiescent solar atmosphere. These observational results are consistent with a situation where the decay in the length scale, driven by the persistent diffusion of the magnetic field to smaller length scales \citep{2011ApJ...730L...3M} has affected the amount of energy supplied to heat the fast wind plasma at its roots, dropping its particle density \citep[e.g.,][]{2008GeoRL..3518103M} and mean Iron charge state. The high FIP of Helium almost certainly ensures its strong sensitivity to a small change in the plasma heating present in the quiescent solar network. Further, this is also consistent with the reduced degree of Iron fractionation measured in, and out of, the ecliptic plane at the same time. While the physical process responsible for the ``FIP effect'' in fast or slow winds is, as yet, undetermined it is clear that the plasma is being less ``aggressively'' heated at its roots - and so, for the fast wind at least, the plasma heating process in the lower solar atmosphere must play an essential role in establishing the degree of fractionation observed \citep{GeissEA95}. It is likely that the excessive low-FIP element fractionation of the slow solar wind is complicated further by the circulation of heated and cooling material \citep[see the cartoon of][]{DePontieu2009} and is far beyond the scope of this Letter in complexity, but an interesting item of further study. Similarly, it is not yet clear how the discrete heating events we discuss here scale in magnitude or number with the complexity (number of small-scale same polarity flux bundles) of the magnetic network vertices, suffice to say that it would appear to play some prominent role in the initial (heating) phase of the mass cycle and is being reflected directly in the particulate emissions of coronal holes than those of the magnetically closed quiet sun and active regions.

The fact that there is still a solar wind at speeds greater than 500km/s is likely further evidence in support of the notion that the fast solar wind is a result of a two stage process, the first to heat the plasma low in the solar atmosphere, combined with a second phase acting away from the solar surface to accelerate the wind to its final measured speed as was speculated by \citet{Parker1991} and more recently revisited by \citet{McIntosh2010} and \citet{McIntosh2011b}. This result is accentuated by a study of the Helium abundance during the space age using the NASA OMNI database which clearly shows a slowly decaying amount of Helium being driven into the heliosphere over the course of the several solar cycles indicative of a slowly decaying background magnetic field and resulting reduction in energy heat supply and resulting mass release.

\section{Interpretation}
We propose that the observed decreases in \ahe{} and the \dfe{} (approaching values expected of the photosphere) in the fast wind indicate a significant change in the process loading material into the fast solar wind during the recent solar minimum. The mean fast solar wind speed shows a small decrease between the two minima \citep[][]{2008GeoRL..3518103M}, but the abundance change observed is profound - indicating to us that the amount of energy deposited into the plasma has changed dramatically. We believe that this is justification for our premise that the process heating the fast solar wind plasma and that which subsequently accelerates it into interplanetary space are, at best, loosely coupled (having different length scales of dissipation) even though they are rooted in the (prevalent) network length scale. As we have stated above the mechanism heating the plasma in unipolar flux concentrations such as those forming the network vertices it is still not clear \citep[][]{DePontieu2011}, we can speculate on why the fast wind \ahe{}, \dfe{} and \qfe{} changed in the last solar minimum, and may be continuing to reduce from minimum to minimum over recent decades. 

Based on the primary inference of \citet{2011ApJ...730L...3M}, the ongoing diffusion of the quiet sun magnetism to smaller length scales reduced the number (and possibly also the strength) of the magnetic elements (e.g., G-Band bright points) in the network vertices. Given that the strongest heating events predominantly occur in the proximity of these unipolar magnetic structures \citep[][]{DePontieu2009}, it would seem sensible to infer that the heating rate depends critically on the arrangement of the discrete flux elements of the vertex and not only their strength\footnote{It is not clear how the heating event frequency and strength scale with the number of flux elements or their strength, but is an important avenue to be tested observationally and numerically \citep[e.g.,][]{2011ApJ...736....9M} in the near future.}. Helium, having the highest first ionization potential will feel this (subtle) change in plasma heating most and Iron, given that it is rapidly ionized, will have less time to establish strong fractionation and a lower charge state. Therefore, we believe that that reduced plasma heating strength (and/or frequency) in the lower solar atmosphere - driven by the diffusion of the field - reduced the ``complexity" of the network vertices, this modified the plasma heating which then led to the changes in fast solar wind composition. We note that it is a different challenge to identify what is happening to the underlying magnetism at the root of the declining Helium abundance over the past several solar cycles.

\begin{acknowledgements}
SWM is supported by NSF ATM-0541567 (ATM-0925177), NASA grants NNG06GC89G, NNX08AL22G, and NNX08AH45G. K$^3$ visited HAO  in the summer of 2011 supported by the NSF Research Experience for Undergraduates (REU) program \-- we greatly appreciate the organization and commitment of the CU/LASP REU program administration group. In addition, we thank Joe Gurman for valuable comments, the \uly{}/SWICS and \ace{}/SWICS instrument teams for the effort taken to distribute and document the ion composition they serve to the community, likewise to the OMNI database staff for their continued effort to preserve an essential physical record. NCAR is sponsored by the National Science Foundation. 
\end{acknowledgements}

\end{document}